\documentclass[12pt,prd]{revtex4-2}
\usepackage{amsmath}
\usepackage{graphicx}

\begin{document}
\title{ Improved semiclassical model for real time evaporation of Matrix black holes. }
\author{David Berenstein, Yueshu Guan}

\affiliation{Department of Physics, University of California, Santa Barbara, CA 93106}
%\date{April 2021}
\begin{abstract}
We study real time classical matrix mechanics of a simplified $2\times 2$ matrix model inspired by the black hole evaporation problem. This is a  step towards making a quantitative  model of real time evaporation of a black hole, which is realized as a bound state of D0-branes in string theory. The model we study is  the reduction of Yang Mills in $2+1$ dimension to $0+1$ dimensions, which has been 
corrected with an additional potential that can be interpreted as a zero point energy for fermions. Our goal is 
to understand the lifetime of such a classical bound state object in the classical regime. To do so, we pay particular attention to when D-particles separate to check that the ``off diagonal modes" of the matrices become adiabatic and use that information to improve on existing models of evaporation.
It turns out that the naive expectation value of the lifetime with the fermionic correction is infinite. This is a logarithmic divergence that arises from very large excursions in the separation between the branes near the threshold for classical evaporation.
The adiabatic behavior lets us get some analytic control of the dynamics in this regime to get this estimate. This divergence is cutoff in the quantum theory due to quantization of the adiabatic parameter,  resulting in a long lifetime of the bound state, with a parametric dependence of order $\log(1/\hbar)$. 
\end{abstract}

\maketitle

\section{Introduction}

Ever since the seminal work of Strominger and Vafa \cite{Strominger:1996sh}, it has been understood not only that black holes in string theory can be built from D-brane bound states, but that in supersymmetric cases they account correctly for the entropy of the black holes. The area law can be justified more generally in holographic duals of field theories by appealing to thermal physics of conformal field theories  \cite{Susskind:1998dq}.
Another feature of quantum black holes is that they can evaporate via Hawking radiation \cite{Hawking:1974sw}. It is expected that this  thermal radiation only stops when the black hole has evaporated completely. One usually models such a system with black body radiation from the horizon. If we have a microscopic description of the black hole, we should be able to do better. We should be able to derive this phenomenon from first principles.
What we are missing is a quantitative model of  real time black hole evaporation. In the quantum theory, the real time physics analysis  becomes prohibitively difficult because of the large numbers of degrees of freedom of matrix models. Also, Monte Carlo simulations will suffer from a severe sign problem. 

On the other hand, the classical physics of these matrix dynamical systems is both chaotic and thermalizes quickly  \cite{Asplund:2011qj, Asplund:2012tg}. The Lyapunov spectrum is consistent with a continuous distribution \cite{Gur-Ari:2015rcq,Hanada:2017xrv}. This classical dynamics regime has served as a starting point for the discussion of real time dynamics in these systems even if it is far from the ideal quantum regime.

Direct Monte-Carlo simulations of matrix quantum mechanics are effective at  getting the thermodynamics of the black hole \cite{Catterall:2008yz,Anagnostopoulos:2007fw}
( see also the more recent  \cite{Hanada:2013rga,Kadoh:2015mka,Filev:2015hia,Berkowitz:2016jlq} and  references therein), and it matches the 10D black hole gravity geometry in \cite{Aharony:1999ti}. The model itself is the Euclidean simulation of the BFSS matrix quantum mechanics \cite{Banks:1996vh}.
These are Euclidean simulations of the partition function. As such, they do not contain real time physics.  To get a handle on the real time evaporation one can study the real time physics of the matrix model classically, rather than the full quantum dynamics. The hope is to find 
a quantitative approach to understanding the black hole evaporation dynamics, which can in principle be compared to the dual gravity picture.  One should study this in the same regime where one  finds fast thermalization and chaos. Once one understands the ingredients in the simplest models one can hope to scale the problem to a full computation in systems of $N\times N$ matrices.  

In this paper we make some partial progress in this direction. The main idea is to use a simple black hole model that arises from a bound state of only two D-particles, with a minimal number of matrices,  and to study in detail how and, more precisely, when the two D-particles separate to form independent particles. 
Our model is in spirit a close cousin of the statistical model proposed in \cite{Berkowitz:2016znt,Berkowitz:2016muc}, where some qualitative aspects of the evaporation can be understood. Their model is studied with various ad-hoc approximations for when a D-particle leaves the bound state. The main assumption being that once one reaches a particular distance away from the black hole, the particle has evaporated away. Our main goal is to improve this description by studying the real time evolution of the system directly rather than relying on the distance between the branes as a proxy for the evaporation. In principle, the improved model would let us study the evaporation in more general contexts where not only one has a black hole, but one can change the classical charges like angular momentum, or the energy, so that one can compare the evaporation physics quantitatively between different configurations. 

The main model we construct is a slight modification of the two matrix model studied in \cite{Berenstein:2016zgj}. In that  model, one can show that the dynamics is generically chaotic and that there are classical flat directions. Unfortunately, in the classical theory these flat directions are of measure zero, so such a black hole would not evaporate: the trajectories might go far in these directions, but they always come back. We need instead an additional (quantum) potential that ensures that once the particles are sufficiently far, that they will keep on separating forever. Rather than put a hard cutoff on the distance, we can use another ingredient that is common in  matrix black holes: they usually arise from supersymmetric constructions. In such setups there are fermionic degrees of freedom as well as bosons. The zero point energy of the fermions depends on the separation of the branes and becomes more negative as they separate further, providing an effective repulsion at long distances. We modify the two matrix model by such a long distance potential where we quench such fermions to their ground state and treat this zero point energy as an additional contribution to the dynamics. Since we are treating the bosons fully  dynamically and classically, this additional zero point energy correction opens up the possibility of classical evaporation in real time. We are then able evolve the dynamics in real time and to compute the statistics of the evaporation events.

In the full dynamics, when branes separate enough, the quantized off-diagonal modes connecting the branes are interpreted as strings stretching between the branes \cite{Witten:1995im}.  The occupation number of these strings stretched between branes can become an adiabatic invariant, which is an important ingredient in studying D-brane scattering problems  \cite{Bachas:1995kx,Douglas:1996yp}. The separation at which this adiabatic behavior takes place   depends on the precise value of the relative velocity between the branes.  After this occupation number is seeded, the dynamics bringing the branes back together is controlled by this invariant. In the classical theory, this is an adiabatic invariant of some modes that become frozen for part of the trajectory and can become dynamical again when the branes get together again. Our computations track this invariant instead of only the distance between the branes. It is only when this invariant is below a particular threshold that the zero-point energy of the fermions can win and the configuration evaporates. Long excursions are still possible and not cut-off automatically. We can evaluate their time of 
return reliably and distinguish them from evaporating solutions.

\section{The setup}

The starting setup is given by the Lagrangian
\begin{equation}
{\cal L}= \frac 12\hbox{Tr} \left((D_t X_1)^2+(D_t X_2)^2 + \frac 12[X_1,X_2]^2\right)
\end{equation}
where $X_1,X_2$ are matrices in the adjoint of $SU(2)$. We use the parametrization in terms of Pauli matrices, so that $X_1$ can be thought of as a three vector $\vec x_1$, whose components are derived from
$X_1= (x_{11} \sigma^1+x_{21} \sigma^2+x_{31} \sigma^3)/\sqrt 2$, and similarly for $X_2$. This is a standard reduction of Yang Mills in $2+1$ dimensions to $0+1$ dimensions as is typical for D-brane effective theories. As discussed in \cite{Berenstein:2016zgj}, in the $A_0=0$ gauge, and in the presence of the gauge constraints, we can use the residual gauge transformations to set $x_{31}=x_{32}=0$ for all time and parametrize the system in terms of a single $2\times 2$ matrix 
\begin{equation}
    U= \begin{pmatrix}x_{11}& x_{12}\\
    x_{21}& x_{22}\end{pmatrix} = \frac{1}{\sqrt2}\begin{pmatrix}\cos(\chi)& -\sin(\chi)\\
    \sin(\chi)& \cos(\chi)\end{pmatrix}\begin{pmatrix}r& r \cos(\theta)\\
    0&r\sin(\theta) \end{pmatrix}\begin{pmatrix}\cos(\phi)& -\sin(\phi)\\
    \sin(\phi)&  \cos(\phi)\end{pmatrix}
\end{equation}
where we have introduced a parametrization in terms of angular coordinates on the right. The $\chi$ angle is a variable that is part of the original gauge transformations, so its canonical conjugate will vanish. The $\phi$ variable is dynamical and it produces $SO(2)$ rotations between the matrices $X_1, X_2$.  Given the parametrization, one can check that $|\vec x_1|^2+|\vec x_2|^2= \hbox{Tr}(X_1^2+X_2^2)=r^2$, so that $r$ is a notion of the radius of the matrix configuration. 

One can also check that $\det(U)^2=-Tr( [X_1,X_2]^2)$, so that $\det(U)= r^2\sin(\theta)=0$ implies that the two matrices commute. Also, this is the potential term of the model. The locus $\theta=0$ is exactly the locus where the two matrices are simultaneously diagonalizable. In a simple example where only $X_1$ is different from zero, and it is diagonal, we get that 
\begin{equation}
X_1= \begin{pmatrix}x_1/2 &0\\
0&-x_1/2\end{pmatrix}.
\end{equation}
In this case the separation between the eigenvalues is $x_1$, and the radius squared is $r^2=x_1^2/2$, so that the radius can be related to the naive notion of separation of the branes as $x_1=\sqrt 2 r$. Along this special flat direction, $\sqrt 2 r$ is interpreted directly as the length of the separation between two D-branes, which are given by the eigenvalues of the matrices $X_1,X_2$ \cite{Witten:1995im}. The angle  $\theta$ therefore parametrizes the extent to which the two matrices are off-diagonal with respect to each other.

If the energy is fixed, when $r$ is very large, $\theta$ is necessarily small. In this regime string theory would dictate that the dynamics is the same as two D-branes with maybe a few strings stretching between them. These strings lead to off-diagonal configurations with small off-diagonal entries. Small fluctuations in $\theta$ at large separations should therefore be associated to these string excitations. The mass or frequency of oscillation of these strings is linear in $r$, and at small velocities for the D-brane centers we have that $\dot \omega/\omega^2\simeq v/r^2 \to 0$ as $r\to \infty$. Thus, when the D-branes are far from each other, these off-diagonal modes are adiabatic. In this far away region the number of strings stretching between the branes is conserved. This is the same reasoning that results in the Born-Oppenheimer approximation becoming invalid when this condition is violated \cite{Douglas:1996yp}, in a region which was labeled the ``stadium". Separating the branes further leads to a linear potential in $r$ proportional to the number of strings. Only if the number of strings exactly vanish can the branes separate forever. In a full quantum theory, this number, when measured far-away,  is an integer. In our case, or in the classical theory, this number is a continuous variable, and the set where it vanishes is of measure zero. This means that in a naive classical theory the branes can never separate indefinitely. There might be long excursions, but they always return after a long time.
This is also seen in a simpler model with only two degrees of freedom
\begin{equation}
    H= \frac12(p_x^2+p_y^2)+ g x^2 y^2
\end{equation}
which results from a consistent truncation of Yang Mills theory \cite{Sav,Aref'eva:1997es}

Such systems do not evaporate, so we would say we have an eternal matrix black hole. We need to fix this first before we calculate further. To do so, it turns out that in supersymmetric setups there are also fermions behaving similarly to the bosons. They are necessary for separation and ``evaporation" of the branes in the quantum theory. Without them, there is a residual zero point energy contribution to the off-diagonal degrees of freedom that also grows linearly in distance and that makes a naive quantum theory with only bosonic degrees of freedom stable. In our classical model we will add a negative contribution that grows linearly in $r$ that we can interpret as the fermionic zero point energy.

When we go to the Hamiltonian description of our modified classical theory we use the following Hamiltonian:
\begin{equation}
    H= \frac 12 p_r^2+\frac 2{r^2} p_\theta^2+\frac{p_\phi^2}{2 r^2 \cos(\theta)^2}+\frac 14 r^4 \sin^2(\theta)-a r\label{eq:Ham}
\end{equation}
This is found both in \cite{Fatollahi:2015fna,Berenstein:2016zgj} and the new ingredient is the addition of the negative term $-ar$.
Here we should interpret the additional constant $a$ as if it is $\hbar$, meaning it should be taken to be small. This term gives a linear repulsive potential at large distances which can guarantee that the branes may be able to separate forever in the classical theory, so long as the number of strings between them is small enough. In that sense, this modified theory can produce classical decay. It is this simple model with the repulsive potential that we study in this paper. This modification cures the eternal matrix black hole problem.

\section{Large excursions and adiabatic invariants.}

As described in the previous section, when the radius variable gets large enough, we expect the variable $\theta$ to become small. 
In that case, we would approximate the Hamiltonian for $\theta$ with small fluctuations around $\theta\simeq 0$. We  write
\begin{equation}
H_\theta \simeq \frac{p_\theta^2}{2 m(t)}+\frac 12 m(t) \omega^2(t) \theta^2(t)
\end{equation}
where $m(t)$ would depend on the variable $r(t)$ and $\omega(t)$ is an effective frequency of the degree of freedom $\theta$. 
It is easy to check that when $r$ is large we have that $\omega(t) \rightarrow \sqrt 2 r(t)$. The adiabatic invariant $N(t)$ is the energy stored in $\theta$ divided by the angular frequency $\omega(t)$:
\begin{equation}
    N(t) = H_\theta(t)/\omega(t)
\end{equation}

The term that depends on angular momentum $p_\phi$ in the total  Hamiltonian given by equation \eqref{eq:Ham} becomes unimportant when $r$ gets very large, but it might correct the dynamics of $r$. It is a type of centrifugal force in the presence of angular momentum $p_\phi$.
We keep it however at finite $r$ in the computation of the adiabatic invariant as a function of time.

Since this effective frequency for $\theta$ becomes large at large $r$, the differential equations associated to the evolution of $\theta$ become stiff. This regime is inevitable: when the D-particles evaporate from the bound state the quantity $r$ will grow without bound.
We checked the adiabatic behavior in $\theta$ and verified that the adiabatic invariant will control the evolution of the variable $r$. Thus, we can cutoff the evolution at an appropriate large $r$, and when we reach it we reflect $p_r\to -p_r$. This will preserve the adiabatic invariant and the energy, giving a trajectory consistent with the main evolution of the system. If we do so, we will miss an accumulated phase for the conjugate angle to the adiabatic invariant. This angle is expect to vary a lot from one escape trajectory to the next, so it is reasonable to expect that it is random. If that is the case, the fact that we chose one value in the reflection should not matter. By the time the mode becomes non-adiabatic again a large phase will be accumulated anyhow. 

\begin{figure}[ht]
\includegraphics[width=10 cm]{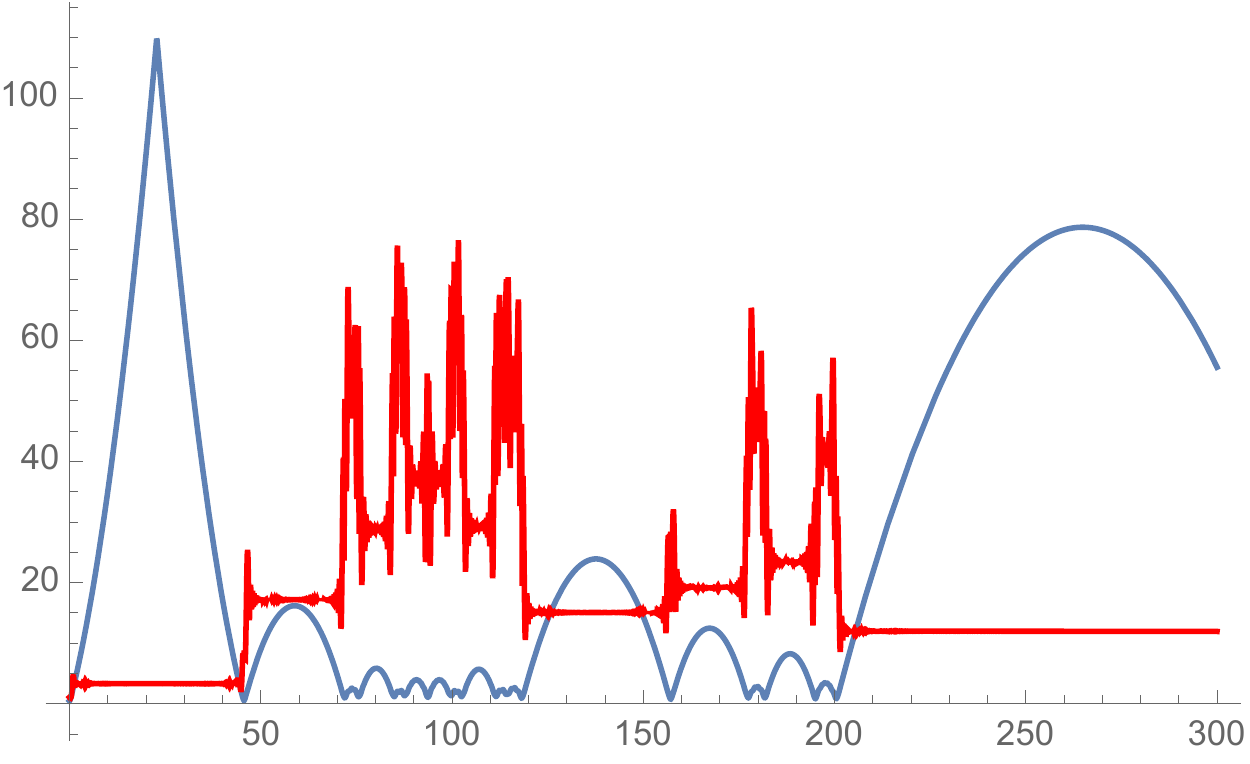}\caption{In blue we have $r(t)$, and in red the numerical value of the adiabatic invariant as a function of time (in arbitrary units). We see that $N(t)$ has various flat steps whenever $r$ is large enough. The initial conditions have $a=0.3, p_\phi= 1.05, E_{tot}= 3$. Notice that the large $N$ is,  the fastest that the trajectory bounces back. 
We use reflections in $r$ at $r=110$, where the adiabatic approximation seems very good. }\label{fig:advsr}. 
\end{figure}

In the adiabatic regime, we expect that the effective Hamiltonian for $r$ becomes
\begin{equation}
    H_{eff}(r)= \frac{p_r^2}{2}+\frac{p_\phi^2}{2 r^2} - a r + N(t)  \omega(r)
\end{equation}
and that $\omega(r) \to \sqrt 2 r$ at large $r$. The variable $N(t)$ is approximately constant in this large $r$ regime. So long as $\sqrt{2} N(t)>a$, the net potential is attractive and linear in the separation at long distance. This is the string intuition for separated D-branes connected by a finite number of strings equal to $N(t)$. An example of these calculations can be seen in figure \ref{fig:advsr}. One can clearly see that $N(t)$ has regions where it varies very little, and then there are jumps that occur when $r$ gets small enough.

\begin{figure}
     \centering
     \includegraphics[width=10 cm]{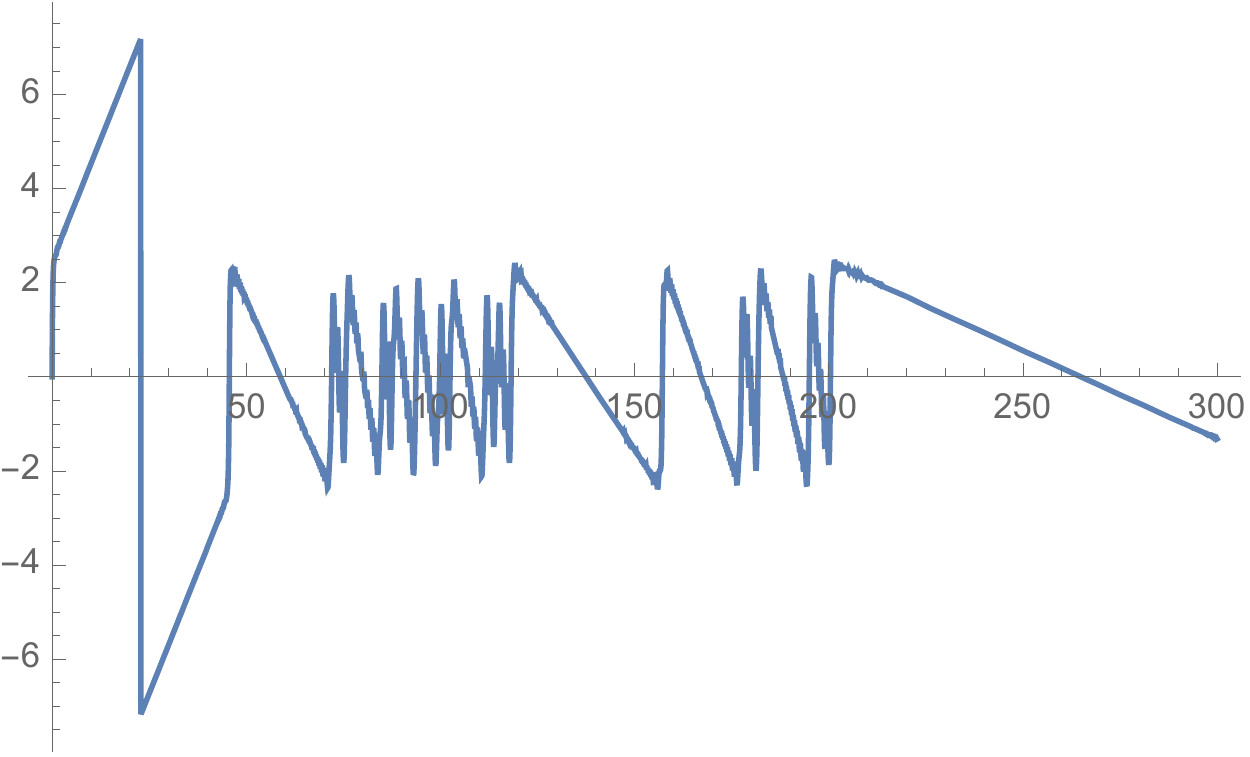}
     \caption{Plot of $ p_r(t)$ vs. $t$. }
     \label{fig:prvst}
 \end{figure}

 We clearly see that for the initial conditions we have studied in figure \ref{fig:advsr},  when $r>20$ the adiabatic invariant can be treated as constant. Next we need to check that in the regime where $\omega(r)\simeq \sqrt 2 r$, for each $N$ we have $\dot p_r$ being approximately constant. When the configuration is attractive, we have $\dot p_r<0$, whereas if the adiabatic mode is below the value of $a$, the configuration is repulsive. 
 This information is depicted in Figure.\ref{fig:prvst}, which shows examples of both repulsive and attractive excursions.
We see clearly that $p_r(t)$ has a linear behavior in time in the adiabatic regimes. We also see one reflection where the momentum was growing  and jumps to a negative value. The reader can verify that reflection in figure \ref{fig:advsr} as well. 
Basically, we can clearly distinguish evaporating solutions to the equations of motion from non-evaporating solutions.

Our goal now is to get the expectation value of the matrix black hole's lifetime. We measure the time from one evaporating event to the next. We measure time between formation (from an escaped trajectory that we bounce back) to the evaporation. To have proper cutoff in time, we need to choose a radius where the system is adiabatic and we can understand that an evaporation event has occurred. Our cutoff in distance is set at $r=20$. Large excursions beyond $r=20$ that return are allowed in the computation of the lifetime, but if they evaporate they are not allowed. 

There are two regimes of interest. The large excursion regime where a trajectory goes beyond the cutoff and returns, and the small radius region (the ``stadium" in the language of \cite{Douglas:1996yp}). Our expectation value of the time to evaporation would have two components. One that measures the time on long excursions beyond the cutoff, and another time that measures time in the stadium.
The long excursion time is expected to have a probability distribution of $N$, which we call $\mu(N)$  which needs to be convoluted with the time of flight of the corresponding large radius excursion, where we can evaluate the time of flight analytically.
\begin{equation}
   \tau_L=\langle \tau_{ex}\rangle = \int^{N_{max}}_{N_{min}} dN \mu(N) \tau(N)
\end{equation}
where $N_{min}= a/\sqrt{2}$ is the minimal value of the adiabatic invariant, whereas $N_{max}$ is determined by the cutoff radius of the long excursions.  The subscript $L$ stands for long. 
There is a second contribution of the stadium where the trajectory is bouncing at low $r$, $\tau_{S}$. The subscript $S$ is both for short excursions and stadium. Once a trajectory leaves the stadium, there is a total probability $p$ that it is evaporated, and $1-p$ that it it is not evaporated. Instead of evaporating, it goes into a long trajectory and returns to the stadium.
The long excursion times can be computed from the distribution $\mu(N)$ that can be measured in the data. 

Both $\tau_S$ and $\tau_L$  depend on at what radius we choose the cutoff in the radius that separates them.
For $\tau_S$ we are measuring times from infall (production) to escape at a fixed radial distance. When we move this radius we have that $\tau_S$ changes: some of the "long excursions" become small excursions. Also, there is an extra time of flight in an evaporation event from one choice to the next.
Similarly, $\tau_L$ varies. The cutoff of $N_{max}$ is changed, and the time of flight is also for a shorter trajectory.
Together they should conspire to produce a final answer for the lifetime that is free of ambiguities. 
We make this explicit assumption before we proceed further.  At this stage we can treat any  dependence on the cutoff as a systematic error of the theory of how to find a classical lifetime. 
We also need to take into account multiple "rescattering" events: where after a return of a long excursion we might have multiple such excursions. 

We now have the stadium time and the long excursion time as two independent quantities, which we treat as statistically independent.
After the first pass in the stadium, with probability $p$ we escape (and then the time is just the time of the stadium), but with probability $(1-p)$ we have a long excursion, which comes back to the stadium. In this case we add a second copy of the time in the stadium. We can keep track of all of these long excursions, so that if have a $k-th$ excursion , it occurs  with decreasing probability $(1-p)^k$. The total time can be summed from a geometric series  as follows 
\begin{eqnarray}
    \langle \tau \rangle &= &\tau_S+(1-p)(\tau_S+\tau_L)+(1-p)^2(\tau_S+\tau_L)+\dots\\
    &=&\tau_S +\frac{(1-p)}{p} (\tau_S+\tau_L)\label{eq:tottime}
\end{eqnarray}
Measuring the statistics of $\tau_S$ and the probability distribution $\mu(N)$, plus checking that the time of flight matches the analytic computation of the time with the simplified effective dynamics  will give us what we need. 
The classical probability of escape is $p= \int_{0}^{N\_{min}} \mu(N) dN $ and is measured by measuring the distribution of $N$ for long excursions. Notice that $p$ varies not because of the end-points in $N$ in the integral determining $p$, but from the total normalization of the distribution $\mu$.

Basically, we have a procedure to extract a notion of a lifetime of the  classical ``matrix bound state". In our data, every time we have an escaping trajectory, the reflection process that we use to deal with the stiffness problem of the differential equation produces a new formation event that can be interpreted as a new collision between a pair of D-branes.

The parts of long returning trajectories that occur beyond the reflection point, where the differential equation is stiff,  need an added time of flight: we fit $\dot p$ for each of these and we also measure $N$, to find the distribution of $N$. We then explicitly check the linear relation between $\dot p$ and $N$. The missing time for very long returning trajectories can be evaluated from the jump in $p$ at the reflection point. 
We explain the implications of this data in the next section.

\section{Classical divergence of the bound state lifetime and quantum finite lifetime.}\label{sec:div}

\begin{figure}
    \centering
    \includegraphics[width=10cm]{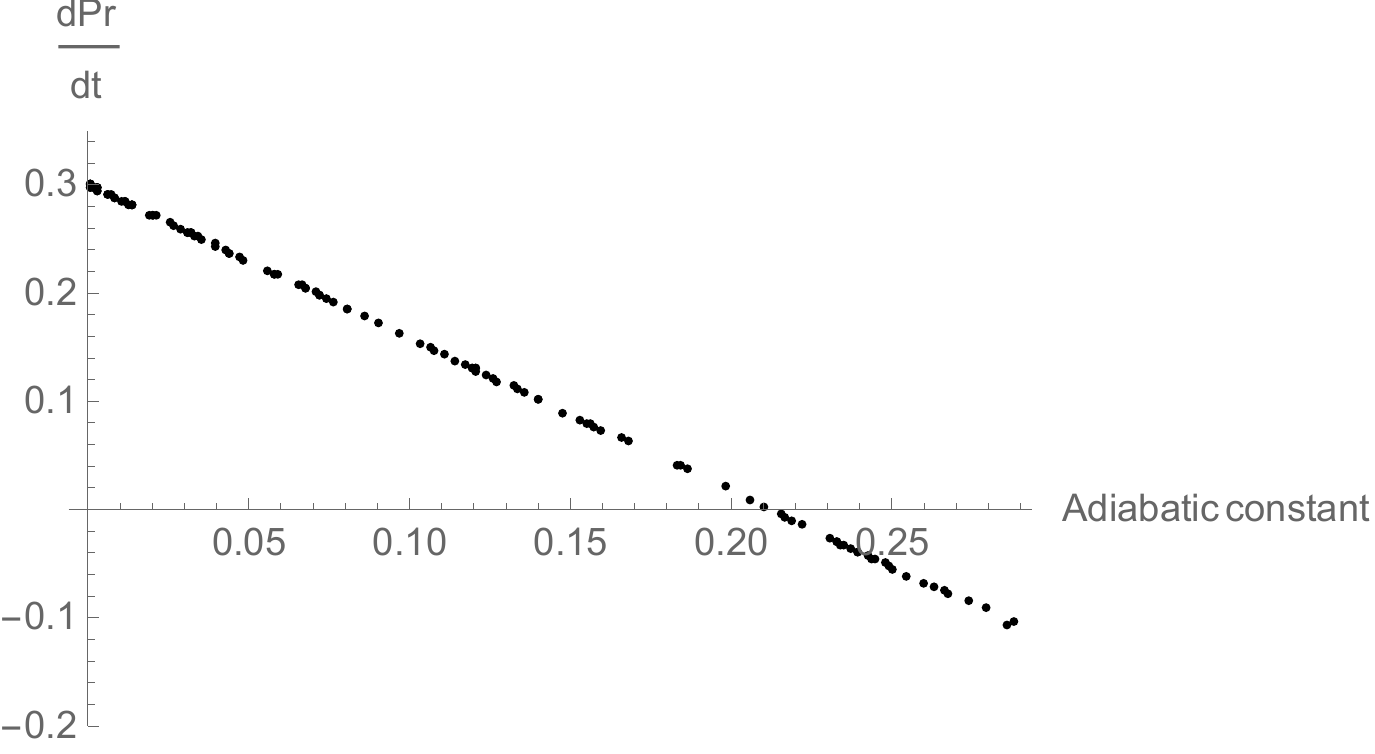}
    \caption{Data of $\dot p_r$ versus the adiabatic invariant extracted from  the dynamics. A fit to the data shows that 
    $\dot p_r \sim 0.2993 -1.41334 N$, very close to the theory $\dot p_r= 0.3-\sqrt 2 N$ (our simulations in this paper are done with $a=0.3$). We have data both above and below 
    the $\dot p_r=0$ line. }
    \label{fig:dpvsad}
\end{figure}

It turns out that the most important contribution to the lifetime of the matrix bound state is the time $\tau_L$.
The argument for this is the following. When we are in the adiabatic regime, the simplified equation of motion of $p_r$ is very simple. 
Essentially
\begin{equation}
    \dot p = a- \sqrt 2 N 
\end{equation}
We can check that the parameters match the data, as in the figure \ref{fig:dpvsad}.

The dynamics in the adiabatic regime is a constant acceleration. Essentially, this part of the trajectory is a parabola. 
We go from $p$ to $-p$ in a time given by
\begin{equation}
t_N = \frac{2 p}{  \sqrt 2 N-a }\label{eq:simple}
\end{equation}
and in this process we return back to the same $r$ due to conservation of energy.
At the cutoff surface we have that 
\begin{equation}
    \frac{p_r^2}{2}+ (\sqrt2N-a)r_{cutoff}= E .
\end{equation}

\begin{figure}
    \centering
    \includegraphics[width=10cm]{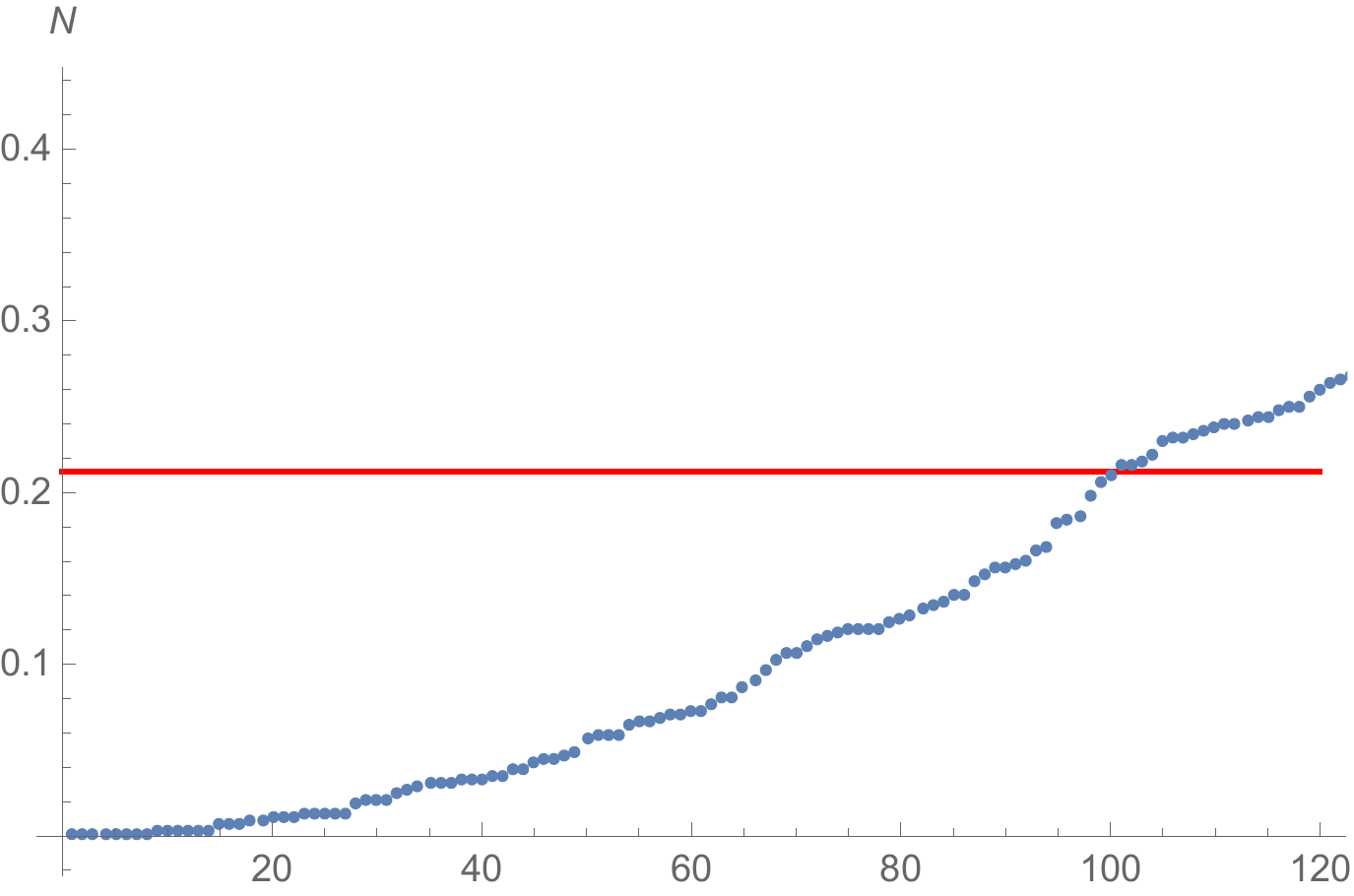}
    \caption{Data for all the $N$, for a total run time of $T_{tot}= 5000$. The critical value of $N$ is shown. The level density is given by the inverse of the slope in the above plot. That slope is finite near the critical value for escape $N_{min}$, indicating that $\mu(N)$ does not vanish near $N_{min}$.}
    \label{fig:ordern}
\end{figure}

What we see is that $p$ is a continuous function of $N$. At the critical escape point for evaporation, when $\sqrt 2 N=a$ we are away from $p_r=0$. 
Near this value the numerator in equation \eqref{eq:simple} is finite and the denominator goes to zero. This means that there are in principle very long excursions in the $r$ direction that come back. We still need to convolve this time with the distribution $\mu(N)$. That is, we have an integral with a logarithmic singularity arising from the lower endpoint of integration $N_{min}$. 
\begin{equation}
     \tau_L=  t_0 \int_{N_{min}} dN \mu(N) \frac1 {N-N_{min}}
\end{equation}
From our data, shown in figure \ref{fig:ordern} the distribution of $N$ is smooth near $N_{min}$, so the classical value of $\tau_L\to \infty$. This divergence appears linearly
in the equation
\eqref{eq:tottime}, so in the end calculation the expectation value of the classical lifetime of the D-brane bound state is infinite.

Our model for evaporation has produced decays in finite time, but it  has not cured the infinite lifetime problem as an expectation value. It is obvious that the divergence
is related to how we introduced the constant $a$, which we claimed was due to quantum effects of fermions. The quantum effects of bosons, and in particular the boson $\theta$ will fix this: it produces a discrete spectrum for $N$ rather than a continuous spectrum.
This quantization of $N$, which is a number operator, is in units of $\hbar$. The value $N_{min}$ is actually forbidden, and the gap to the next occupied state is $a$. The value of $a$ is a fermion zero point energy that will cancel the boson zero point energy in a supersymmetric setup. The $-1/2 \hbar \omega= - a \omega $ is cancelled by $\frac 12 \hbar \omega$ for the off-diagonal bosons. The occupation number $N$ is an integer copy of
$2a$. We find this way that a lower value of $N$ for returning trajectories  is $2 N_{min}$, rather than $N_{min}$. The logarithmic divergence gets replaced by a large number, which parametrically scales as
\begin{equation}
    \tau_L \sim \log(1/\hbar) \sim \tau_0\log(N_{max}/N_{min})
\end{equation}
where $\tau_0$ is a typical return time for intermediate value of $N$ between $N_{min}$ and $N_{max}$. Notice that the value of $N_{max}$ in the cutoff radius is determined by $p_r=0$ which leads to zero time of return. 

The introduction of $\hbar$ as a lower cutoff for $N$ will render the end calculation finite. 
However, this is beyond the classical analysis where $N$ is continuous rather than discrete. Implementing the calculation in detail with a semiclassical analysis requires making sense of the continuous variable $N$ of the classical theory. At this level, it requires a model of how to handle that variable from the classical to the quantum theory. Notice that in our sample data the cutoff of $N_{max}<2 N_{min}$ is below this naive quantum threshold for long excursions, so the modeling issue becomes crucial if progress is to be made. 

This might be special to this matrix model of $2\times 2$ matrices. In larger matrices we would have a similar expression
\begin{equation}
   \tau_L=  t_0 \int_{N_{min}} d^k N \mu(N) \frac1 {N_{tot}-N_{min}} \label{eq:ktimesk}
\end{equation}
where $k$ counts the total number of off-diagonal modes and $N_{tot}$ is the total adiabatic invariant. So long as an analog of $N_{min}$ appears in the right hand side of \eqref{eq:ktimesk} there will be a logarithmic divergence. If this correction to the models is not included, the logarithmic correction vanishes because the measure removes the divergence. This avoidance of the singularity would be similar to entropic effects removing certain divergences in multi matrix models \cite{Krauth:1998xh}. This question can only be resolved if one has a better understanding of the full quantum theory and not just this semiclassical approach. Within our setup with the fermions quenched and producing a net repulsion at long distances, the logarithmic divergence is there, even if the precise number might be somewhat suppressed due to measure effects.

\subsection{Systematics}

We now have to tackle the systematic uncertainties that are introduced in our modeling. 
Let us begin with the main model itself. The fermionic zero
point energy that we added modifies the dynamics slightly with respect to the naive matrix model we would originally have. 
It does produce the main effect we needed for classical evaporation, but we still need additional quantum effects to get a finite lifetime. 

Here, it is important to understand that in the end we are doing quantum physics. We can argue that at the radius for adiabatic 
behavior that we have used for a cutoff, the value of $N$ should be quantized. We can take this radius as a location where the system itself measures the adiabatic invariant, and then we can evolve the dynamics based on the quantized $N$ rather than a continuous $N$. 

There are many possible models. We can, for example, take the integer $N$ that is closest to the continuous value of $N$. If we do so, we need a prescription for how to handle the classical dynamics after this point. If we want to conserve energy, a jump in $N$ can be done by rescaling $\theta$ and $\dot \theta$ simultaneously by the same scaling factor to get to the correct $N$. At the same time, we modify $p_r$ but leave $r$ fixed. This can be done classically, but it is somewhat arbitrary.
In situations with large $N$, this would only modify the dynamics of $p_r$ slightly and leads to a reasonable estimate where we replace the integral over $N$ by a sum with a lower cutoff.

The probability $p$ is then computed and gives us some value. In our case, because our numerical experiments are below threshold for the full value of $\hat N$ that is needed at the cutoff, we would get that $p=1$ and the lifetime would be only the time in the stadium. This seems a bit contrived, but it can match a naive escape problem where escaping means reaching a particular radius. This is actually the philosophy of the crude model of \cite{Berkowitz:2016znt}. In our case, we are adding time of flight dynamics. The important difference is that when reaching a radius, we can get there with different velocities. If we have bigger matrices, this radius might be too far to be meaningful, or too small to be adiabatic. A better choice is to let the effective cutoff be dictated dynamically by the onset of adiabaticity itself and the distribution of $N$ that should arise there.  

A different prescription is to assume that the state for the angular $\theta$ variable is a coherent state with $\langle \hat N \rangle=N$. This has classical minimal uncertainty and is the closest analog to a classical initial condition.
We can then read the probability distribution of the eigenvalues of $\hat N$ from this prescription. 
This has a smoother behavior in the distribution of $\hat N$ and would lead to having to convolve an additional probability distribution of values of $N$ with the classical $\mu(N)$ we already have. The probability $p$ would then also receive contributions from higher values of $N$. We would also need to forbid measurements of states that have very high values of $\hat N$: they would not reach the cutoff radius and might lead to a small negative time contribution to $\tau_L$. If this is very small anyhow, it does not matter much.

A third possibility is to measure the probability distribution of the eigenvalues of $\hat N$ using a thermal distribution for $\hat N$
such that the expectation value of $N$ is known. This one will have more variance in $N$, but can in principle also be justified arguing some sort of thermalization with other degrees of freedom is taking place and a lack of further knowledge of the quantum system of the $\theta$ variable is at play. 

Neither of these is satisfactory as a semiclassical add-on in the case of our data set: we do not have enough trajectories with high 
enough values of $N$ that reach the desired point. For very high energy D-brane bound states, all of these should give somewhat similar answers. In our case, not only will the answers be distinct, but the quantum effects are too pronounced to trust the other part of the 
classical computation. To choose between these options, we need a fuller quantum computation.  

There are additional issues on systematics. Our definition of the time $\langle \tau \rangle$ depends on the cutoff: we measure the time of arrival at this location. A better way to address this is to compare trajectories where $\theta$ is identically zero at all times, but where we cutoff the radius at different locations. We can compare them to each other at different radii and we would realize that we need to subtract the time of flight between them. This way we should get at an answer that is in principle independent of the choice of radial cutoff. Since in principle we need to vary the quantity $p_\phi$, we need to compare to the point of closest approach to $r=0$ that is allowed. This would be the impact parameter of the exiting particle (and incoming particle). Basically, we would need to compare the time delay relative to the case of no interactions between the D-branes. This is in principle better behaved and one can consider the relation between a time delay and a phase shift as a starting point for a more quantum description of the dynamics.

\section{Discussion}

In this paper we have seen that it is possible to enhance the classical dynamics of matrix models that arise from string theory ideas to get configurations that mimic black hole evaporation. Basically, a bound state of D-branes can evaporate classically.
Our results improve on previous models by keeping track of the the number of open strings between the D-branes in the dynamics. The number of these strings are counted in the classical theory by an adiabatic invariant when D-branes are well separated from each other. This allows us to better distinguish configurations that evaporate from those that do not. 
As a starting point, this allows us in principle to compare different bound states with different energies and total angular momentum.
The improved follow up of the dynamics can be checked numerically, as we showed: we are able to get a  distribution of adiabatic invariants and to directly see the adiabatic portions of the trajectories in  data.  In our case, the choice of variables we used to express this dynamics makes the splitting between radial separation and contributions to the energy from non-trivial 
commutators natural.

When we consider bound states of more D-branes, this simple separation of variables is not possible in this way. We would need instead to follow up some matrix version of distances between configurations. For example, the idea of a large eigenvalue
for a radius squared matrix variable might do what we need \cite{Berkowitz:2016znt}. If we are able to follow up the adiabatic behavior at this large radius, we can do a better analysis of the evaporation problem for bound states of more D-branes and compare different configurations to each other. 

When we looked more closely into the trajectories that are near the threshold of evaporation, we found that they could lead to very long excursions that collectively add to a logarithmic divergence of the bound state's expectation value of the lifetime. This divergence is classical and is removed by quantum effects, leading to a large log enhancement of the lifetime of the black hole. Understanding the quantum corrections is crucial. At this stage, this leads to a large systematic error in the regime of most interest: where quantum corrections are important. Basically, the classical analysis we have done on its own is not enough to provide reliable quantitative data in this regime. Qualitatively, we have improved the description of evaporation events by removing the introduction of an ad-hoc cutoff in the radius. Instead, once a radius is chosen, we can check adiabaticity and in principle follow trajectories past this point.
We can then in principle remove the choice of the radius, as described in section \ref{sec:div}. 
A natural location for this cutoff would be some sort of dynamical radius of a black hole, which can be obtained from Euclidean Monte-Carlo simulations \cite{Hanada:2008gy}. More precisely, we expect that the adiabatic behavior becomes important at some finite distance from the horizon rather than at the horizon itself. 

Our model uses fermion degrees of freedom that have been quenched to their ground state. The spectrum of these fermions in matrix configurations has been studied in \cite{Berenstein:2013tya}. There is a region where these become gapless, which also suggests a radius for black holes.
The radius of cutoff needs to be beyond this point as well. When considering various values of $N$, a naive analysis also suggests that
the larger the $N$, the harder it is to evaporate the D-brane bound state. On the other hand, many fermions are actually heavy in this limit and should be quenched to their ground state when a D-particle is exiting this gapless region already. This should also be done with the corresponding off-diagonal heavy bosons, except that it is hard to distinguish these from the diagonal modes. To take into account these effects, the upshot should be that the distribution of the number of strings connecting the escaping D-brane to the black hole should not scale like $N$ in the quantum limit. A full calculation needs an improved description of this separation of variables. If such a separation of variables is possible, it has been argued that one can also 
get the black hole thermodynamics directly, by taking into account the non-trivial residual interactions between the  eigenvalues interpreted as D-branes \cite{Wiseman:2013cda}.

For us, even at small values of the number of D-branes, the black holes are longer lived than a naive dimensional analysis would suggest.
This depends critically on how the femrions are supposed to be treated in the full theory. In a different treatment, we argued that certain measure effects might make the effect go away. Until a direct check on the quantum dynamics is performed, we do not have a sufficiently firm answer to this question. 
There are other situations where logarithms arise in black hole problems. Namely, the Hayden-Preskill information retrieval problem \cite{Hayden:2007cs}. Naively, the two questions asked seem very different, but the fact that both give rise to logarithmic corrections to questions about time delays leads us to speculate that there is a possibly deep connection between these two phenomena.

At lower energies and for different numbers of matrices in the matrix quantum mechanical models, the existence or absence of bound states at threshold can also become important in computing the details of a scattering event. This bound state at threshold only occurs for dimension reductions of SYM in the 10D model, as computed by the index \cite{Moore:1998et}.

\acknowledgments

We are grateful to M. Hanada for very useful correspondence. Work of D.B. supported in part by the Department of Energy under grant DE-SC 0011702.


\begin{thebibliography}{99}


%\cite{Strominger:1996sh}
\bibitem{Strominger:1996sh}
A.~Strominger and C.~Vafa,
``Microscopic origin of the Bekenstein-Hawking entropy,''
Phys. Lett. B \textbf{379}, 99-104 (1996)
doi:10.1016/0370-2693(96)00345-0
[arXiv:hep-th/9601029 [hep-th]].
%2857 citations counted in INSPIRE as of 15 Apr 2021

%\cite{Susskind:1998dq}
\bibitem{Susskind:1998dq}
L.~Susskind and E.~Witten,
``The Holographic bound in anti-de Sitter space,''
[arXiv:hep-th/9805114 [hep-th]].
%820 citations counted in INSPIRE as of 15 Apr 2021


%\cite{Hawking:1974sw}
\bibitem{Hawking:1974sw}
S.~W.~Hawking,
``Particle Creation by Black Holes,''
Commun. Math. Phys. \textbf{43}, 199-220 (1975)
[erratum: Commun. Math. Phys. \textbf{46}, 206 (1976)]
doi:10.1007/BF02345020
%8847 citations counted in INSPIRE as of 15 Apr 2021

%\cite{Asplund:2011qj}
\bibitem{Asplund:2011qj}
C.~Asplund, D.~Berenstein and D.~Trancanelli,
``Evidence for fast thermalization in the plane-wave matrix model,''
Phys. Rev. Lett. \textbf{107}, 171602 (2011)
doi:10.1103/PhysRevLett.107.171602
[arXiv:1104.5469 [hep-th]].
%68 citations counted in INSPIRE as of 16 Apr 2021

%\cite{Asplund:2012tg}
\bibitem{Asplund:2012tg}
C.~T.~Asplund, D.~Berenstein and E.~Dzienkowski,
``Large N classical dynamics of holographic matrix models,''
Phys. Rev. D \textbf{87}, no.8, 084044 (2013)
doi:10.1103/PhysRevD.87.084044
[arXiv:1211.3425 [hep-th]].
%47 citations counted in INSPIRE as of 16 Apr 2021

%\cite{Gur-Ari:2015rcq}
\bibitem{Gur-Ari:2015rcq}
G.~Gur-Ari, M.~Hanada and S.~H.~Shenker,
``Chaos in Classical D0-Brane Mechanics,''
JHEP \textbf{02}, 091 (2016)
doi:10.1007/JHEP02(2016)091
[arXiv:1512.00019 [hep-th]].
%61 citations counted in INSPIRE as of 22 Apr 2021
Copy to ClipboardDownload


%\cite{Hanada:2017xrv}
\bibitem{Hanada:2017xrv}
M.~Hanada, H.~Shimada and M.~Tezuka,
``Universality in Chaos: Lyapunov Spectrum and Random Matrix Theory,''
Phys. Rev. E \textbf{97}, no.2, 022224 (2018)
doi:10.1103/PhysRevE.97.022224
[arXiv:1702.06935 [hep-th]].
%13 citations counted in INSPIRE as of 22 Apr 2021


%\cite{Catterall:2008yz}
\bibitem{Catterall:2008yz}
S.~Catterall and T.~Wiseman,
``Black hole thermodynamics from simulations of lattice Yang-Mills theory,''
Phys. Rev. D \textbf{78}, 041502 (2008)
doi:10.1103/PhysRevD.78.041502
[arXiv:0803.4273 [hep-th]].
%143 citations counted in INSPIRE as of 26 Apr 2021

%\cite{Anagnostopoulos:2007fw}
\bibitem{Anagnostopoulos:2007fw}
K.~N.~Anagnostopoulos, M.~Hanada, J.~Nishimura and S.~Takeuchi,
``Monte Carlo studies of supersymmetric matrix quantum mechanics with sixteen supercharges at finite temperature,''
Phys. Rev. Lett. \textbf{100}, 021601 (2008)
doi:10.1103/PhysRevLett.100.021601
[arXiv:0707.4454 [hep-th]].
%181 citations counted in INSPIRE as of 26 Apr 2021








%\cite{Hanada:2013rga}
\bibitem{Hanada:2013rga}
M.~Hanada, Y.~Hyakutake, G.~Ishiki and J.~Nishimura,
``Holographic description of quantum black hole on a computer,''
Science \textbf{344}, 882-885 (2014)
doi:10.1126/science.1250122
[arXiv:1311.5607 [hep-th]].
%85 citations counted in INSPIRE as of 16 Apr 2021

%\cite{Kadoh:2015mka}
\bibitem{Kadoh:2015mka}
D.~Kadoh and S.~Kamata,
``Gauge/gravity duality and lattice simulations of one dimensional SYM with sixteen supercharges,''
[arXiv:1503.08499 [hep-lat]].
%35 citations counted in INSPIRE as of 07 May 2021

%\cite{Filev:2015hia}
\bibitem{Filev:2015hia}
V.~G.~Filev and D.~O'Connor,
``The BFSS model on the lattice,''
JHEP \textbf{05}, 167 (2016)
doi:10.1007/JHEP05(2016)167
[arXiv:1506.01366 [hep-th]].
%43 citations counted in INSPIRE as of 16 Apr 2021


%\cite{Berkowitz:2016jlq}
\bibitem{Berkowitz:2016jlq}
E.~Berkowitz, E.~Rinaldi, M.~Hanada, G.~Ishiki, S.~Shimasaki and P.~Vranas,
``Precision lattice test of the gauge/gravity duality at large-$N$,''
Phys. Rev. D \textbf{94}, no.9, 094501 (2016)
doi:10.1103/PhysRevD.94.094501
[arXiv:1606.04951 [hep-lat]].
%59 citations counted in INSPIRE as of 19 Oct 2021



%\cite{Aharony:1999ti}
\bibitem{Aharony:1999ti}
O.~Aharony, S.~S.~Gubser, J.~M.~Maldacena, H.~Ooguri and Y.~Oz,
``Large N field theories, string theory and gravity,''
Phys. Rept. \textbf{323}, 183-386 (2000)
doi:10.1016/S0370-1573(99)00083-6
[arXiv:hep-th/9905111 [hep-th]].
%5009 citations counted in INSPIRE as of 07 May 2021


%\cite{Banks:1996vh}
\bibitem{Banks:1996vh}
T.~Banks, W.~Fischler, S.~H.~Shenker and L.~Susskind,
``M theory as a matrix model: A Conjecture,''
Phys. Rev. D \textbf{55}, 5112-5128 (1997)
doi:10.1103/PhysRevD.55.5112
[arXiv:hep-th/9610043 [hep-th]].
%2889 citations counted in INSPIRE as of 07 May 2021

%\cite{Berkowitz:2016znt}
\bibitem{Berkowitz:2016znt}
E.~Berkowitz, M.~Hanada and J.~Maltz,
``Chaos in Matrix Models and Black Hole Evaporation,''
Phys. Rev. D \textbf{94}, no.12, 126009 (2016)
doi:10.1103/PhysRevD.94.126009
[arXiv:1602.01473 [hep-th]].
%30 citations counted in INSPIRE as of 15 Apr 2021


%\cite{Berkowitz:2016muc}
\bibitem{Berkowitz:2016muc}
E.~Berkowitz, M.~Hanada and J.~Maltz,
``A microscopic description of black hole evaporation via holography,''
Int. J. Mod. Phys. D \textbf{25}, no.12, 1644002 (2016)
doi:10.1142/S0218271816440028
[arXiv:1603.03055 [hep-th]].
%15 citations counted in INSPIRE as of 15 Apr 2021


%\cite{Berenstein:2016zgj}
\bibitem{Berenstein:2016zgj}
D.~Berenstein and D.~Kawai,
``Smallest matrix black hole model in the classical limit,''
Phys. Rev. D \textbf{95}, no.10, 106004 (2017)
doi:10.1103/PhysRevD.95.106004
[arXiv:1608.08972 [hep-th]].
%8 citations counted in INSPIRE as of 15 Apr 2021

%\cite{Witten:1995im}
\bibitem{Witten:1995im}
E.~Witten,
``Bound states of strings and p-branes,''
Nucl. Phys. B \textbf{460}, 335-350 (1996)
doi:10.1016/0550-3213(95)00610-9
[arXiv:hep-th/9510135 [hep-th]].
%1561 citations counted in INSPIRE as of 16 Apr 2021

%\cite{Bachas:1995kx}
\bibitem{Bachas:1995kx}
C.~Bachas,
``D-brane dynamics,''
Phys. Lett. B \textbf{374}, 37-42 (1996)
doi:10.1016/0370-2693(96)00238-9
[arXiv:hep-th/9511043 [hep-th]].
%384 citations counted in INSPIRE as of 22 Apr 2021

%\cite{Douglas:1996yp}
\bibitem{Douglas:1996yp}
M.~R.~Douglas, D.~N.~Kabat, P.~Pouliot and S.~H.~Shenker,
``D-branes and short distances in string theory,''
Nucl. Phys. B \textbf{485}, 85-127 (1997)
doi:10.1016/S0550-3213(96)00619-0
[arXiv:hep-th/9608024 [hep-th]].
%609 citations counted in INSPIRE as of 16 Apr 2021

%\cite{Fatollahi:2015fna}
\bibitem{Fatollahi:2015fna}
A.~H.~Fatollahi,
``Regge Trajectories by 0-Brane Matrix Dynamics,''
J. Geom. Symm. Phys. \textbf{17}, 231-242 (2016)
doi:10.7546/giq-17-2016-231-242
[arXiv:1506.02961 [hep-th]].
%2 citations counted in INSPIRE as of 22 Apr 2021

  \bibitem{Sav}
  G. Z.~ Baseyan, S. G.~ Matinyan and G. K. ~ Savvidi, JETP Lett. {\bf 29}, 585 (1979)
  
 %\cite{Aref'eva:1997es}
\bibitem{Aref'eva:1997es} 
  I.~Y.~.Aref'eva, P.~B.~Medvedev, O.~A.~Rytchkov and I.~V.~Volovich,
  ``Chaos in M(atrix) theory,''
  Chaos Solitons Fractals {\bf 10}, 213 (1999)
  [hep-th/9710032].
  %%CITATION = HEP-TH/9710032;%% 


%\cite{Hanada:2008gy}
\bibitem{Hanada:2008gy}
M.~Hanada, A.~Miwa, J.~Nishimura and S.~Takeuchi,
``Schwarzschild radius from Monte Carlo calculation of the Wilson loop in supersymmetric matrix quantum mechanics,''
Phys. Rev. Lett. \textbf{102}, 181602 (2009)
doi:10.1103/PhysRevLett.102.181602
[arXiv:0811.2081 [hep-th]].
%83 citations counted in INSPIRE as of 28 Apr 2021


%\cite{Krauth:1998xh}
\bibitem{Krauth:1998xh}
W.~Krauth, H.~Nicolai and M.~Staudacher,
``Monte Carlo approach to M theory,''
Phys. Lett. B \textbf{431}, 31-41 (1998)
doi:10.1016/S0370-2693(98)00557-7
[arXiv:hep-th/9803117 [hep-th]].
%155 citations counted in INSPIRE as of 07 May 2021


%\cite{Berenstein:2013tya}
\bibitem{Berenstein:2013tya}
D.~Berenstein and E.~Dzienkowski,
``Numerical Evidence for Firewalls,''
[arXiv:1311.1168 [hep-th]].
%19 citations counted in INSPIRE as of 28 Apr 2021


%\cite{Wiseman:2013cda}
\bibitem{Wiseman:2013cda}
T.~Wiseman,
``On black hole thermodynamics from super Yang-Mills,''
JHEP \textbf{07}, 101 (2013)
doi:10.1007/JHEP07(2013)101
[arXiv:1304.3938 [hep-th]].
%20 citations counted in INSPIRE as of 28 Apr 2021


%\cite{Hayden:2007cs}
\bibitem{Hayden:2007cs}
P.~Hayden and J.~Preskill,
``Black holes as mirrors: Quantum information in random subsystems,''
JHEP \textbf{09}, 120 (2007)
doi:10.1088/1126-6708/2007/09/120
[arXiv:0708.4025 [hep-th]].
%589 citations counted in INSPIRE as of 28 Apr 2021

%\cite{Moore:1998et}
\bibitem{Moore:1998et}
G.~W.~Moore, N.~Nekrasov and S.~Shatashvili,
``D particle bound states and generalized instantons,''
Commun. Math. Phys. \textbf{209}, 77-95 (2000)
doi:10.1007/s002200050016
[arXiv:hep-th/9803265 [hep-th]].
%316 citations counted in INSPIRE as of 07 May 2021

\end{thebibliography}
\end{document}